\documentclass[aps,amsmath,amssymb,reprint, double column,prl, showkeys]{revtex4-2}

\usepackage{graphicx,epsfig}
\usepackage{amssymb}
\usepackage{amsmath}
\usepackage{bm}
\usepackage{textcomp}
\usepackage{color}

\usepackage{soul}

\begin{document}
\setcounter{page}{1}

\title[]{Melting phase diagram of bubble phases in high Landau levels}
\author{K. A. \surname{Villegas Rosales}}
\author{S. K. \surname{Singh}}
\author{H. \surname{Deng}}
\author{Y. J. \surname{Chung}}
\author{L. N. \surname{Pfeiffer}}
\author{K. W. \surname{West}}
\author{K. W. \surname{Baldwin}}
\author{M. \surname{Shayegan}}
\affiliation{Department of Electrical Engineering, Princeton University, Princeton, New Jersey 08544, USA}

\date{\today}

\begin{abstract}

A low-disorder, two-dimensional electron system (2DES) subjected to a large perpendicular magnetic field and cooled to very low temperatures provides a rich platform for studies of many-body quantum phases. The magnetic field quenches the electrons' kinetic energy and quantizes the energy into a set of Landau levels, allowing the Coulomb interaction to dominate. In excited Landau levels, the fine interplay between short- and long-range interactions stabilizes bubble phases, Wigner crystals with more than one electron per unit cell. Here we present the screening properties of bubble phases, probed via a simple capacitance technique where the 2DES is placed between a top and a bottom gate and the electric field penetrating through the 2DES is measured. The bubbles formed at very low temperatures screen the electric field poorly as they are pinned by the residual disorder potential, allowing a large electric field to reach the top gate. As the temperature is increased, the penetrating electric field decreases and, surprisingly, exhibits a pronounced minimum at a temperature that appears to coincide with the melting temperature of the bubble phase. We deduce a quantitative phase diagram for the transition from bubble to liquid phases for Landau level filling factors $4\leq\nu\leq5$.

\end{abstract}

\maketitle

Two-dimensional electron systems (2DESs) in a perpendicular magnetic field reveal a fascinating set of many-body quantum phases \cite{Perspectives.Pinczuk.Das.Sarma,Jain.CFbook.2007,Halperin.Fractionalbook.2020}. In the lowest orbital Landau level (LL), there is a plethora of fractional quantum Hall states. In excited LLs, however, the nodes in the wavefunction lead to a weakened short-range interaction and a preferred long-range order, manifested by the formation of charge-density-wave and stripe/nematic phases \cite{Fogler.PRB.1996,Koulakov.PRL.1996,Moessner.PRB.1996,Shibata.PRL.2001,Haldane.PRL.2000,Fradkin.PRB.1999}. Consistent with this expectation, magnetotransport data for very low disorder 2DESs at very low temperatures have revealed anisotropic phases at half-filled, high-index LLs \cite{Lilly.PRL.1999.a,Du.SSC.1999} which are interpreted as stripe (or nematic) phases. Moreover, away from exact half-fillings, e.g., at LL filling factors $\nu\simeq i + 1/4$ and $\simeq i + 3/4$, there are unusual phases with a vanishing longitudinal resistance and a Hall resistance that is quantized at a value corresponding to the nearest integer quantum Hall state (IQHS), namely, at $ih/e^2$ and $(i+1)h/e^2$, respectively ($i$ is an integer $\geq 2$) \cite{Cooper.PRB.1999,Eisenstein.PRL.2002,Deng.PRB.2012,Deng.PRL.2012}. These are believed to be bubble phases, Wigner crystal states with more than one electron per unit cell, as shown in Fig. 1(a).  Because they are pinned by the small but ubiquitous disorder, they are insulating but phenomenologically appear as reentrant IQHSs (RIQHSs). Even though the bulk of the 2DES is insulating, the longitudinal resistance vanishes because of the conducting edge states of the underlying LLs (see Fig. 1(b)). Besides magnetotransport, numerous other experimental techniques have been employed to study the bubble phases; these include measurements of non-linear $I$-$V$ \cite{Cooper.PRB.1999,Wang.PRB.2015,Bennaceur.PRL.2018}, microwave resonance \cite{Lewis.PRL.2002,Lewis.PRL.2004}, and surface acoustic waves \cite{Friess.NatPhys.2017,Friess.PRL.2018}. The results of all these measurements are consistent with the presence of bubble phases.


Here we present experimental data, probing the bubble phases and their melting into a liquid state, using a technique that measures their screening efficiency. As highlighted in Fig. 1(c), this is a simple capacitance technique \cite{Eisenstein.PRL.1992,Eisenstein.PRB.1994,Deng.PRL.2019,Ma.PRL.2020} where the application of an AC voltage between the top and back gates induces an electric field $E_{P}$ that penetrates through the 2DES. The magnitude of $E_{P}$ depends on the screening efficiency of the 2DES bulk, and the size of the penetrating current $I_{P}$ is then probed in response to $E_{P}$. At the lowest temperatures, we find that the pinned bubble phases screen the electric field between the gates poorly and a large $I_{P}$ is observed. With increasing temperature, the screening efficiency of the 2DES improves and $I_{P}$ drops, as expected. However, $I_{P}$ shows a distinct minimum at a temperature that corresponds to the melting temperature of the bubble phases. This observation suggests that the bubble phases becomes particularly efficient at screening near their melting.  We use the data to construct a bubble-liquid melting phase diagram as a function of filling factor.  


\begin{figure*}[t!]
  \centering
    \psfig{file=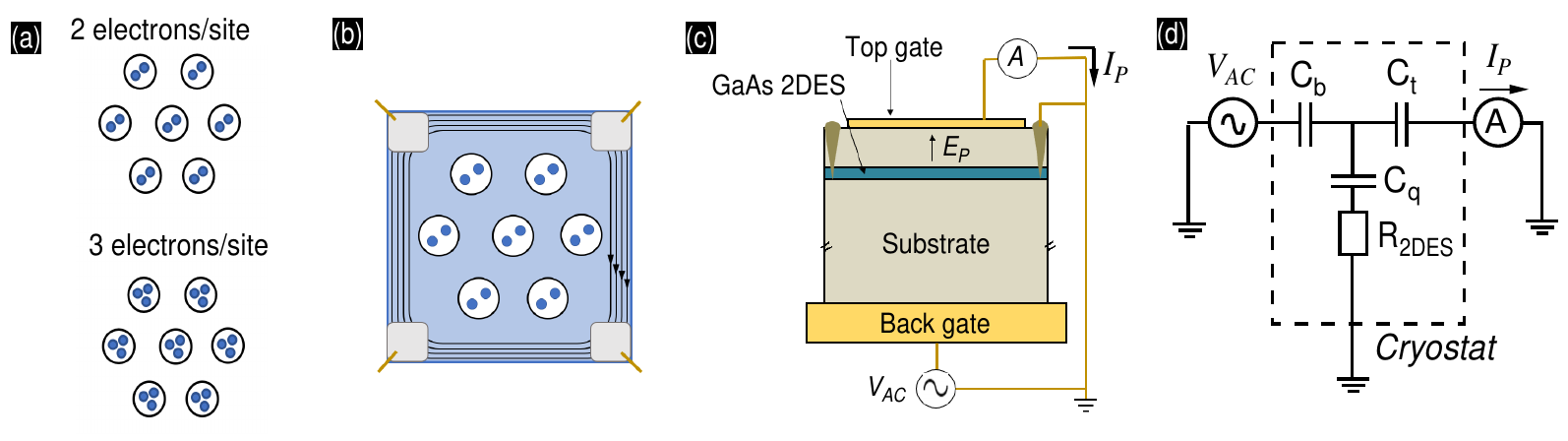, width=1\textwidth}
  \centering
  \caption{\label{transport}
(a) Schematic representations of bubble phases that have 2 or 3 electrons per lattice site. Bubble phases are formed in the excited Landau levels. (b) Measurement geometry (van der Pauw), where the grey squares represent the diffused electrical contacts to the GaAs 2DES. The edge currents flow near the perimeter of the sample, while the bubble phases occupy the bulk of the system. (c) Penetrating electric field measurement set-up. An AC excitation voltage ($V_{AC}$) is applied between the top and back gates and the GaAs 2DES screens the established electric field, allowing only a penetrating electric field ($E_{P}$) to pass through. From the top gate we collect the penetrating current ($I_{P}$) induced by $E_{P}$. (d) A lumped-element circuit model of our device. The different components such as back-gate capacitance ($C_{b})$, top-gate capacitance ($C_{t}$), and the 2DES quantum capacitance ($C_{q}$) and resistance ($R_{2DES}$) are located inside of the cryostat (dashed box).
  }
  \label{fig:transport}
\end{figure*}

\begin{figure}[h!]
  \centering
    \psfig{file=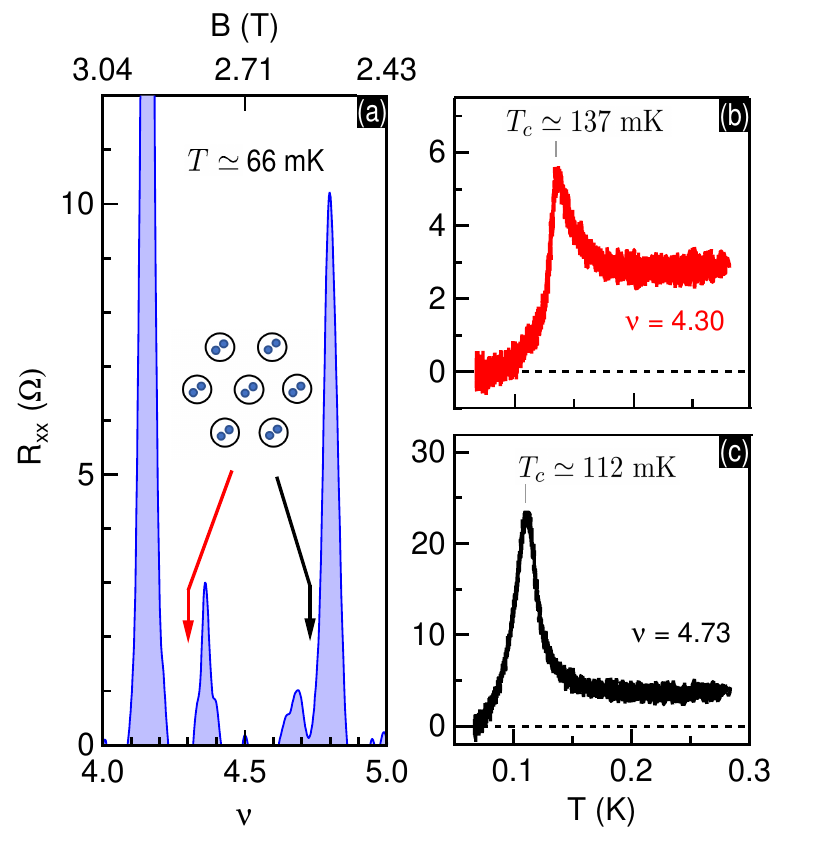, width=0.49\textwidth}
  \centering
  \caption{\label{Ip}
(a) Longitudinal resistance ($R_{xx}$) vs. Landau level filling factor ($\nu$) for our 2DES with density \textit{n} $=2.96\times10^{11}$ cm$^{-2}$, at temperature \textit{T} $\simeq66$ mK. The trace represents the resistance along the ``easy-axis" direction. The red and black arrows point to the regions where the bubble phases, exhibiting a RIQHS behavior, are observed. The inset shows a bubble phase with 2 electrons/site that forms at the lowest temperatures. (b) $R_{xx}$ vs. $T$, measured at $\nu=4.30$ (red) and $4.73$ (black). As the temperature is raised, the resistance peaks at $T_{c}$.
  }
  \label{fig:Ip}
\end{figure}

We studied a 2DES confined to a 30-nm-wide modulation-doped GaAs quantum well grown on a GaAs (001) substrate. The Si dopant atoms were placed in doping wells \cite{Chung.PRM.2020}, leading to a very high-quality 2DES. The sample has density \textit{n} $=2.96\times10^{11}$ cm$^{-2}$ and a low-temperature mobility $\mu\simeq30\times 10^{6}$ cm$^{2}$/Vs. We measured a $4$ mm $\times$ $4$ mm van der Pauw geometry sample with alloyed (InSn) electrical contacts at the corners. The device is mounted on a header using In that serves as a back gate, and on top has a deposited, semi-transparent, 15-nm-thick Al film, top gate. We illuminated our sample before the measurements \cite{footnote1}. We used lock-in techniques for magnetoresistance ($\simeq17$ Hz) and capacitance measurements. For the latter, the frequency (\textit{f}) range was $2$ Hz $\lesssim$ \textit{f} $\lesssim1000$ Hz. The sample and a calibrated, RuO thermometer next to it were placed on the cold finger of a dry dilution refrigerator.


Figure 2(a) provides magnetoresistance data as a function of $\nu$ at the temperature \textit{T} $\simeq66$ mK. Near $\nu=4.3$ and $4.7$ (marked by arrows), the longitudinal resistance ($R_{xx}$) has vanishing values and the Hall trace is quantized to the resistance value of the nearest IQHS, signaling a RIQHS behavior \cite{Lilly.PRL.1999.a,Du.SSC.1999,Cooper.PRB.1999,Lewis.PRL.2002,Deng.PRB.2012,Wang.PRB.2015,Eisenstein.PRL.2002}. See the Supplementary Material SM \cite{SM} for $R_{yy}$ and $R_{xy}$ traces. These characteristics have been associated with disorder-pinned bubble phases, which are observed in the highest-quality 2DESs. Pinned bubble phases have vanishingly small conductivity, but in transport measurements, the insulating behavior is shunted by the underlying edge currents of the filled LLs (see, e.g., Fig. 1(b)). We will return to the temperature dependence of the $R_{xx}$ shown in Figs. 2(b,c) later in the manuscript.

\begin{figure*}[t!]
  \centering
    \psfig{file=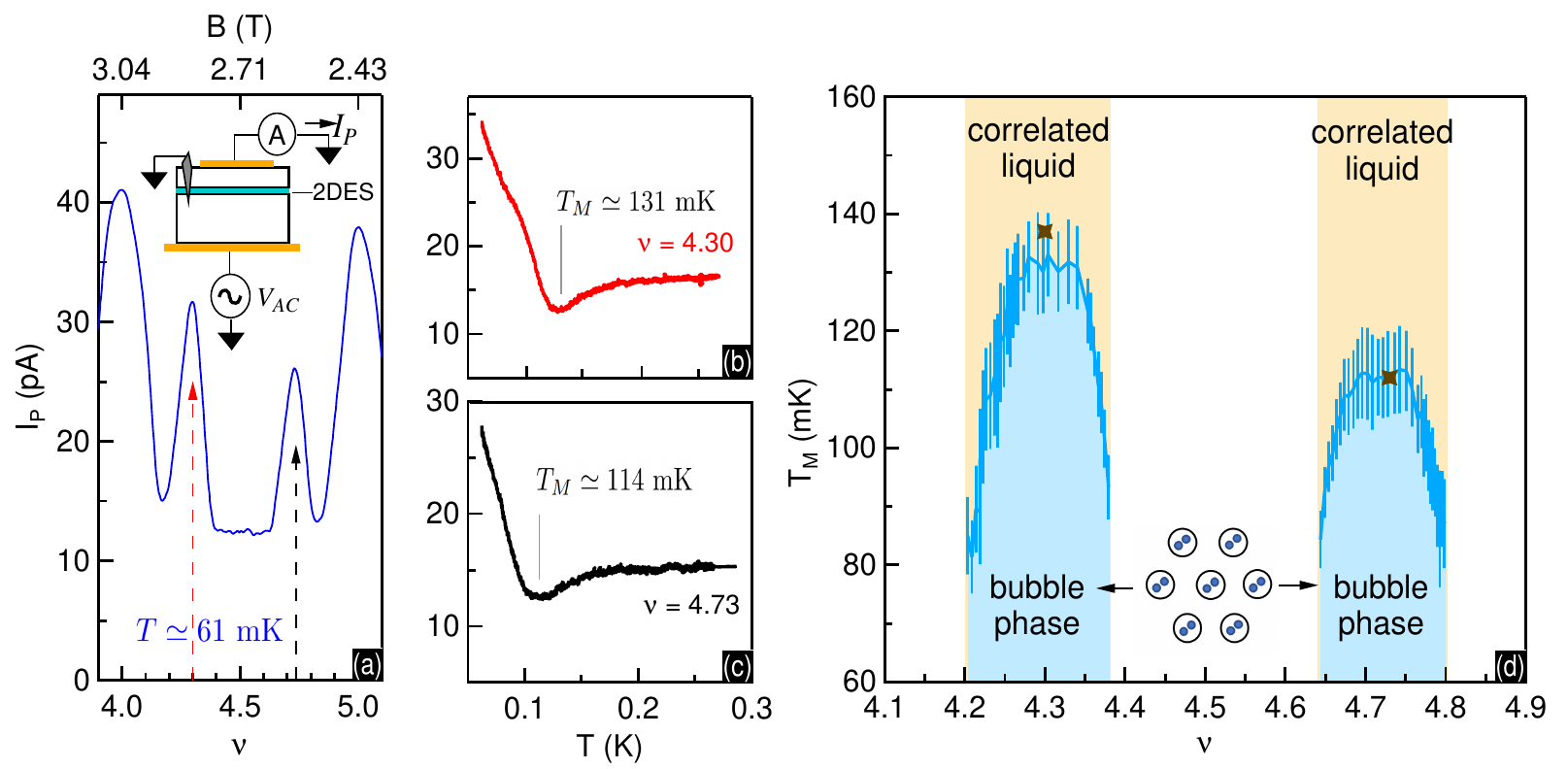, width=1\textwidth}
  \centering
  \caption{\label{Ip}
(a) Penetration current ($I_{P}$) vs. $\nu$ at $T\simeq61$ mK. The bubble phases are marked by dashed lines. The inset shows the capacitance set-up. We apply an AC excitation voltage $V_{AC}$ $=10$ mV between the gates at frequency $f=213$ Hz. (b) Temperature dependence of $I_{P}$ for $\nu=4.30$ and $4.73$. A local minimum in $I_{P}$ is observed at $T_{M}$, which we associate with the melting of the bubble phase. (c) Thermal melting phase diagram for the bubble phases for $4< \nu < 5$. The blue regions are the bubble phases delimited by $T_{M}$. The yellow regions indicate the liquid states at high temperatures. The two brown crosses denote the maxima found in the non-monotonic temperature dependence of $R_{xx}$ in Figs. 2(b,c). The cartoon shows a bubble phase with 2 electrons/site.
}
  \label{fig:Ip}
\end{figure*}

By probing the screening efficiency of the 2DES, we derive information regarding its bulk . Figure 3(a) shows the sample's screening efficiency for $\nu$ ranging from 4 to 5, at $T\simeq61$ mK \cite{footnote5}. At the $\nu=4$ and $5$ IQHSs, $I_{P}$ is a maximum, indicating minimal screening, i.e., an insulating bulk as the quasiparticles are localized by the disorder potential. As $\nu$ increases from $\nu=4$, $I_{P}$ decreases and reaches a minimum at $\nu\simeq4.15$, hinting at the delocalization of quasiparticles that are able to screen better the penetrating electric field. For larger $\nu$, $I_{P}$ rises and peaks at $\nu=4.30$; the local maximum in $I_{P}$ coincides with the bubble phase region in transport experiments, namely the vanishing of $R_{xx}$ (see Fig. 2(a)). In pinned electron solids, the electrons are fixed to their positions and are unable to screen effectively the penetrating electric field, thus leading to signatures of minimal screening (maximal $I_{P}$) \cite{Deng.PRL.2019,Ma.PRL.2020}, as seen in Fig. 3(a) at $\nu\simeq4.30$. When $\nu$ starts at 5 and decreases, $I_{P}$ follows a similar trend, reaching a local maximum at $\nu\simeq4.73$.

In between the bubble phases ($4.40 \lesssim \nu \lesssim 4.65$), $I_{P}$ is minimal. This range of $\nu$ coincides with the region where anisotropic stripe/nematic phases are detected in magnetotransport \cite{Fogler.PRB.1996,Koulakov.PRL.1996,Moessner.PRB.1996,Lilly.PRL.1999.a,Du.SSC.1999,Fradkin.PRB.1999} (see Fig. 2(a)). Stripe/nematic phases are unidirectional charge density waves that are highly conductive along the stripe/nematic direction at finite temperatures, thus an impinging electric field should be screened, in agreement with the local wide $I_{P}$ minimum seen in Fig. 3(a). Note that we observe the same signatures of pinned bubble and stripe/nematic phases in $I_{P}$ at higher LLs, up to $\nu=7$ \cite{SM}.

By measuring $I_{P}$ at fixed $\nu$ as the temperature is raised, we capture the screening behavior of the bubble phases as they melt. Figures 3(b,c) show $I_{P}$ vs. \textit{T} for $\nu=4.30$ and $4.73$. At the lowest temperatures, $I_{P}$ has large values consistent with pinned bubble phases. At high temperatures, $I_{P}$ saturates at a value that is lower than its maximum low-temperature value, consistent with a melted state, i.e., a liquid phase that has a higher screening efficiency than the pinned bubble phase. One would expect that as the bubbles melt, $I_{P}$ would change monotonically between these two limits as the electrons become unpinned with increasing temperature. Surprisingly, however, in between these two limits, $I_{P}$ reaches a local minimum at temperature $T_{M}$, implying maximal screening at this temperature \cite{footnote4}.

We attempt to understand qualitatively the non-monotonic $I_{P}$ response with a lumped-element circuit model as shown in Fig. 1(d) \cite{footnote2}. The 2DES is characterized by its quantum capacitance ($C_{q}$) and bulk resistance ($R_{2DES}$). Since values of the top and bottom capacitances ($C_{t}$ and $C_{b}$) are fixed, changes in $I_{P}$ are a direct consequence of variations in the 2DES impedance $Z_{2DES}=R_{2DES}+\frac{1}{j\omega C_{q}}$. The local minimum in $I_{P}$, at $T_{M}$, implies a dip in $Z_{2DES}$, which could mean that $R_{2DES}$ decreases, or $C_{q}$ increases, or both.

We associate $T_{M}$ with the melting temperature of the bubble phases based on the following considerations. First, a qualitatively similar behavior in $I_{P}$ vs. $T$ traces was recently seen when studying the screening properties of the Wigner crystal at very small $\nu$ (\textit{lowest} LL) in low-density GaAs 2D electron and hole systems \cite{Deng.PRL.2019,Ma.PRL.2020}. Associating the temperature for maximum screening with the melting temperature of Wigner crystal, Ref. \cite{Deng.PRL.2019} found the measured dependence of this temperature on $\nu$ to be consistent with the melting phase diagrams reported previously for the Wigner crystal in GaAs 2DESs \cite{Goldman.PRL.1990,Williams.PRL.1991,Paalanen.PRB.1992,Chen.NatPhys.2006}. Second, theories for the melting of a 2D solid \cite{Kosterlitz.JPC.1973,Halperin.PRL.1978,Nelson.PRB.1979,Young.PRB.1979} predict a divergence in the compressibility near the melting temperature. The compressibility is proportional to $C_{q}$ in Fig. 1(d), and the 2DES screening ability is directly related to $Z_{2DES}$. A large compressibility would lead to increased screening. We surmise that the maximal screening at our measured $T_{M}$ might be related to an increase in $C_{q}$ as the bubble phase melts.

A natural question that arises is whether the non-monotonic behavior of $I_{P}$ and the value of $T_{M}$ are intrinsic to the 2DES and not an artifact of our measurement circuitry. As discussed in the SM \cite{SM} we tested this by measuring $I_{P}$ for frequencies ranging from $21.3$ to $1000$ Hz. The non-monotonic $I_{P}$ behavior and the value of $T_{M}$ are frequency independent over this range.

Associating \textit{T$_{M}$} with the melting temperature, a plot of our measured \textit{T$_{M}$} vs. $\nu$, as shown in Fig. 3(d), provides the bubbles' thermal melting phase diagram. The light-blue and yellow regions represent the bubble phases and correlated liquids, respectively. The error bars give an estimate of the uncertainty in determining the temperature of the local minimum in $I_{P}$ vs. $T$ traces. The left-side bubble-liquid boundary has a relatively broad maximum around $\nu=4.30$, and on its flanks $T_{M}$ decreases quickly reaching values as low as $\simeq 80$ mK at $\nu\simeq4.20$ and $\simeq4.38$. The right-side bubble-liquid boundary also has a broad maximum near $\nu=4.73$ and, away from this maximum, $T_{M}$ falls rapidly to $\simeq80$ mK at $\nu\simeq4.65$ and $4.80$. Note that $80$ mK is our reliable lower temperature limit for detecting a minimum in $I_{P}$. The $\nu$ range of the left-side bubble to liquid boundary in Fig. 3(d) falls within the available theoretical calculations for the bubble phase stability. Density-matrix-renormalization-group \cite{Shibata.PRL.2001} and Hartree-Fock \cite{Cote.PRB.2003,Goerbig.PRB.2004} calculations predict $4.23\leq \nu \leq 4.39$ and $4.21\leq \nu \leq 4.44$, respectively, for the $\nu$ window where the bubbles are stable. In the simplest scenario the regions around $\nu=4.30$ and $4.73$ should be particle-hole symmetric.

The melting of bubble phases has also been studied in transport measurements. Deng \textit{et al.} \cite{Deng.PRL.2012,Deng.PRB.2012} performed resistance measurements in a 2DES of very similar density and mobility to our sample. They showed that in the bubble phases' regime there is a sharp transition in $R_{xy}$ as the temperature decreases. At high temperatures, $R_{xy}$ has the classical Hall resistance value, and as the temperature decreases it quickly transitions to the resistance value of the nearest IQHS. Such a sharp transition was used to define the critical temperature ($T_{c}$) for the formation/melting of bubble phases. The longitudinal resistance that accompanies the sharp transition in $R_{xy}$, has a non monotonic behavior and it peaks at the same $T_{c}$. The measured $T_{c}$ in the $R_{xx}$ temperature dependence can therefore be also used to pinpoint the formation/melting of bubble phases.  Based on their measured $T_{c}$, Deng \textit{et al.} \cite{Deng.PRL.2012} constructed thermal melting phase diagrams for bubble phases between $2\leq\nu\leq4$. Their deduced diagrams also exhibit a dome-shaped boundary, similar to those seen in Fig. 3(d).

It is instructive to compare our results for $I_{P}$ with the temperature dependence of our $R_{xx}$ data. Figures 2(b,c) feature $R_{xx}$ vs. \textit{T} for $\nu=4.30$ and $4.73$. At the lowest temperatures $R_{xx}$ is very close to zero, which is a manifestation of pinned bubble phases: the divergent bulk resistance is shunted by the underlying edge currents of the filled LLs. For \textit{T} $\gtrsim 170$ mK, where the bubble phases have melted and the system is composed of unpinned quasiparticles, $R_{xx}$ reaches a non-zero saturation value. The resistance reaches a maximum at an intermediate temperature $T_{c}$ similar to what has been reported previously in Refs. \cite{Deng.PRL.2012,Deng.PRB.2012,Ro.PRB.2019,Ro.PRB.2020}, as summarized in the preceding paragraph. In our measurements, we find that $T_{c}$ is indeed very close to $T_{M}$ deduced from screening efficiency data (see Figs. 3(b,c)). This can be best seen in the phase diagram in Fig. 3(d) where we have added our measured $T_{c}$ as brown crosses. Note that our measured $T_{c}$ values coincide (to within $5\%$) with those reported by Deng \textit{et al.} \cite{Deng.PRB.2012}.



Further contrasting of our data can be done with microwave resonance results from a very similar GaAs 2DES, again with comparable density and mobility to ours \cite{Lewis.PRL.2002,Lewis.PRL.2004}. At $T\simeq$ 50 mK, Lewis \textit{et al.} \cite{Lewis.PRL.2002} found resonances in the ranges $4.20\lesssim\nu\lesssim4.37$ and $4.62\lesssim\nu\lesssim4.82$, providing support to the picture of pinned bubble phases. These filling factor ranges overlap with those where we see bubble phases in our experiments (Fig. 3(d)). Reference \cite{Lewis.PRL.2002} also reported that the resonances are strongest at low temperatures and disappear when the temperature exceeds $\simeq110$ mK. This is somewhat smaller than our measured $T_{M}\simeq131$ mK. Given our measured $T_{c}$ and $T_{M}$ values, it appears that the melting temperatures derived from screening and transport measurements are slightly larger than those from microwave resonances. 

In summary, we report the first screening efficiency measurements of pinned bubble phases in excited LLs. As the bubble phases melt, a minimum in $I_{P}$ at temperature $T_{M}$ is found, signaling maximal screening. We associate $T_{M}$ with the bubble phase melting temperature and construct a melting phase diagram for the bubbles near $\nu=4.30$ and $4.70$. We would like to highlight that the higher screening at an intermediate temperature appears to be a ubiquitous phenomenon for the melting of magnetic-field-induced electron solids formed in 2D systems, be it in the lowest \cite{Deng.PRL.2019,Ma.PRL.2020} or the excited LLs, thus it begs a rigorous theoretical explanation.


\begin{acknowledgments}

We acknowledge support by the National Science Foundation (NSF) Grant DMR 2104771 for measurements. For sample synthesis and characterization, we acknowledge support by NSF Grants ECCS 1906253 and MRSEC DMR 1420541, and the Gordon and Betty Moore Foundation's EPiQS Initiative (Grant No. GBMF9615 to L.N.P.). We thank David Huse for illuminating discussions.

\end{acknowledgments}


\begin{thebibliography}{99}
\bibitem{footnote1} At 15 K, we shine light with a red-LED (next to the sample) for 30 minutes, and then proceed to mK temperatures.
\bibitem{footnote5} The reported $I_{P}$, which serves as a proxy for the screening efficiency, comes after calibrating the signal to remove the impact of parasitic capacitances. It is the out-of-phase (capacitive) component of the penetrating current.
\bibitem{footnote4} In the SM we report $I_{P}$ vs. $\nu$ at different fixed temperatures, also displaying the melting of bubble phases \cite{SM}.
\bibitem{footnote2} Note that the parasitic capacitances and resistances (of, e.g., wires and contacts) have not been included.
\end{thebibliography}

\end{document}


\setcounter{page}{1}

\title[]{Supplemental Material for ``Melting phase diagram of bubble phases in high Landau levels"}
\author{K. A. \surname{Villegas Rosales}}
\author{S. K. \surname{Singh}}
\author{H. \surname{Deng}}
\author{Y. J. \surname{Chung}}
\author{L. N. \surname{Pfeiffer}}
\author{K. W. \surname{West}}
\author{K. W. \surname{Baldwin}}
\author{M. \surname{Shayegan}}
\affiliation{Department of Electrical Engineering, Princeton University, Princeton, New Jersey 08544, USA}
\date{\today}

\maketitle   

\section{Additional magnetoresistance data}

Figures S1(a-c) show additional magnetoresistance ($R_{xx}$, $R_{yy}$ and $R_{xy}$) data for $\nu$ ranging from $4$ to $5$. The anisotropic state occurs at $\nu\simeq4.5$, and it can be seen in Figs. S1(a,b), where the blue $R_{xx}$ trace in Fig. S1(a) shows vanishingly small values, while the grey $R_{yy}$ trace in Fig. S1(b) displays a peak of $\simeq650$ $\Omega$. In the anisotropic phase regime, the easy-axis direction is defined as the measurement direction which shows an $R_{xx}$ that has very small values as in the blue trace in Fig. S1(a). The hard-axis direction is perpendicular to the easy-axis direction; see the trace in Fig. S1(b). The bubble phases are marked in all three different traces. Deep minima in $R_{xx}$ and $R_{yy}$ with resistance values very close to zero resistance are seen in both Figs. S1(a,b). The Hall trace (Fig. S1(c)) shows the reentrant behavior to the nearest integer quantum Hall state at the positions where the arrows point to. The Hall resistance around $\nu\simeq4.50$ has not been averaged with the negative magnetic field trace. The averaging procedure, usually used when the 2DES has anisotropies, such as in the stripe/nematic region, tends to give a classical Hall line near $\nu\simeq4.50$ \cite{Sajoto.PRL.1993,Goldman.PRL.1993,Chung.NatMat.2021,Deng.PRB.2012}.

\begin{figure*}[h!]
  \centering
    \psfig{file=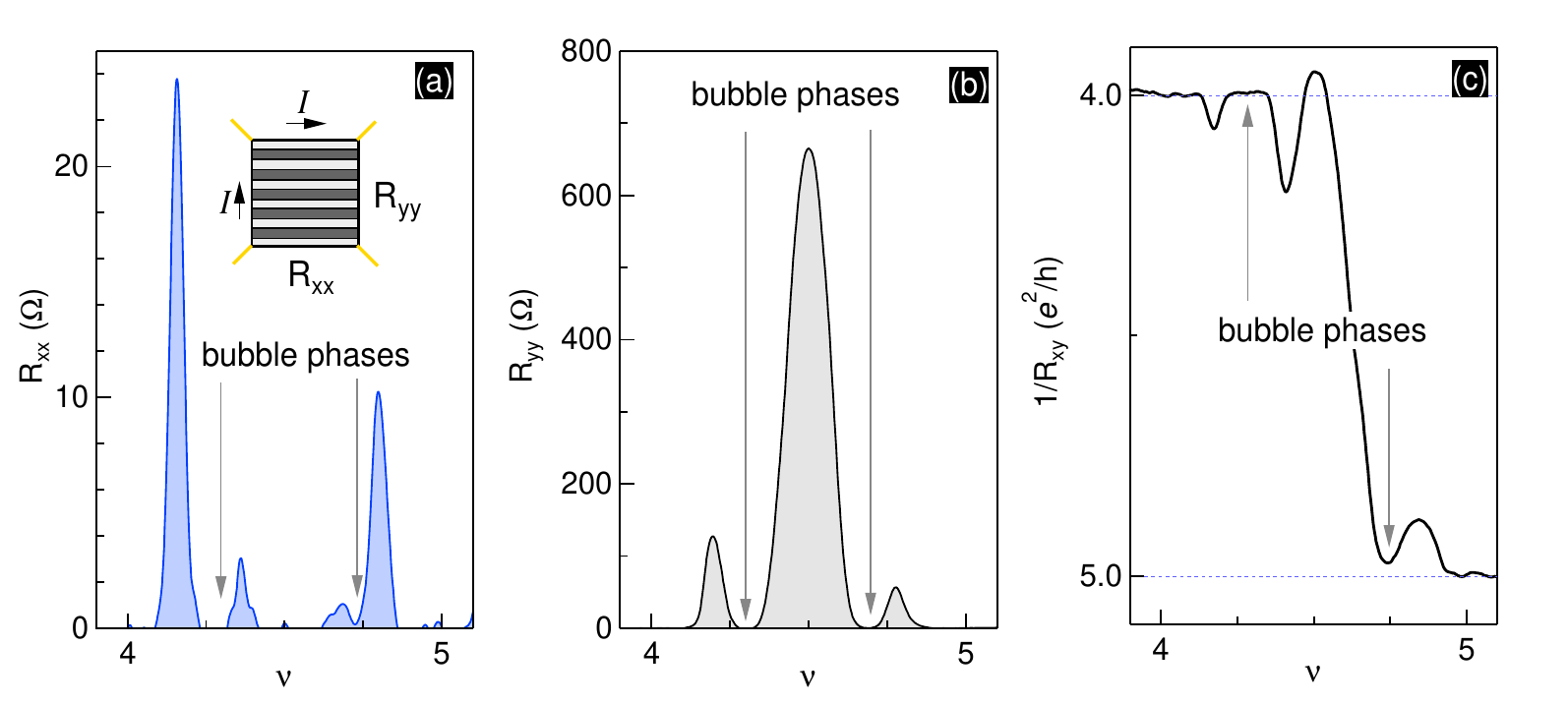, width=1\textwidth}
  \centering
  \caption{\label{transport} 
   (a) Longitudinal magnetoresistance ($R_{xx}$) vs. Landau level filling factor ($\nu$) along the easy-axis direction. (b) $R_{yy}$ vs. $\nu$ along the hard-axis direction. The inset shows the stripe/nematic phase. $R_{xx}$ and $R_{yy}$ are seen next to the measurement geometry. (c) Inverse of the Hall resistance ($1/R_{xy}$) vs. $\nu$. The arrows point to the bubble phases regions and their reentrant behavior to the nearest integer quantum Hall state.
  }
  \label{fig:transport}
\end{figure*}

\clearpage

\section{Signal processing and calibration}

The lumped-element circuit shown in Fig. S2(a) is a simplification of our measurement device; the parasitic capacitances and resistances of wires have not been included. The circuit permits us to qualitatively understand the penetrating current ($I_{P}$) measurements as a function of the 2DES parameters: its resistance ($R_{2DES}$) and quantum capacitance ($C_{q}$). The top- and back-gate capacitances are denoted with $C_{b}$ and $C_{t}$, respectively. At zero magnetic field, we estimate, from geometry considerations, $C_{b}$ and $C_{t}$ to be $\simeq3.7$ pF and $\simeq3.7$ nF, respectively. To estimate the quantum capacitance we use $C_{q}=A \frac{m_{b} e^{2}}{\pi \hbar^{2}}\simeq717$ nF \cite{Serge.APL.1988}, where $m_{b}=0.067m_{0}$ is the GaAs band mass, and $A\simeq16$ mm$^{2}$ is the area of the sample. An excitation voltage ($V_{AC}$) is applied to the back gate and $I_{P}$ is measured from the top gate. 

In Fig. S2(b), we show the calculated $I_{P}$ as a function of $R_{2DES}$ for a frequency $f=213$ Hz, with $C_{q}\simeq717$ nF and $V_{AC}=0.01$ V. The in-phase (resistive) and out-of-phase (capacitance) components are labeled. The traces show $I_{P}$ as a function of the 2DES impedance, from small to large values. For very small $R_{2DES}$, the in- and out-of-phase signals are very small. At $R_{2DES}\simeq10$ $k\Omega$, both signals start to increase. When $R_{2DES}$ is close to $\simeq500$ $k\Omega$, the out-of-phase signal increases rapidly, and the in-phase signal exhibits a peak. Increasing $R_{2DES}$ further causes the out-of-phase signal to saturate at $\simeq50$ pA, while the in-phase signal approaches zero.  The infinite $R_{2DES}$ limit is of interest for the calibration and separation of the device's signal from other spurious components. A 2DES in a strong integer quantum Hall state (IQHS) has a vanishingly small conductivity, i.e., nearly infinite $R_{2DES}$. Experimentally, once an IQHS is identified, the device's in-phase signal should be approximately zero, and the out-of-phase part should tend to $50$ pA. Hence, it is possible to correct for the phase and offset errors coming from the parasitic components.

Based on the qualitative behavior of $I_{P}$ discussed above, we now interpret our measured data as a function of perpendicular magnetic field ($B$). Figure S3 shows the measured $I_{P}$ vs. $B$ at four different frequencies $2.13$, $21.3$, $213$, and $1065$ Hz. When the 2DES is in a strong IQHS, e.g., at Landau level filling factors $\nu=1$, $2$, and $3$, $I_{P}$ is maximal. The infinite $R_{2DES}$ in an IQHS leads to a maximum in $I_{P}$. Meanwhile, the in-phase signal has non-zero values.  Note that $I_{P}$ (out-out-phase) is in the nA range, in contrast to the pA calculated in Fig. S2. The twisted pair wires used to electrically access the top and back gates have a parasitic capacitance that shunts the device's capacitance.

To gain access to the 2DES $I_{P}$, without the spurious parasitic signal, we proceed to calibrate the signal as follows. First, we calculate the phase angle ($\theta$) between the real and imaginary parts at $\nu=2$. The in-phase component ought to be zero, thus $I_{P}$ is rotated by the phase angel $\theta$. As a result, the rotated in-phase current signal is close to zero, as shown in Fig. S4(b). Next, we remove the offset at $B=0$ with the well-grounded assumption that $I_{P}$ tends to zero because the 2DES screens the penetrating electric field very well. The out-of-phase signal after the phase rotation and offset correction can be seen in Fig. S4(a). We now notice that $I_{P}$ ranges between zero and $40$ pA. All the data presented in the main manuscript come after carrying out this calibration procedure.

\begin{figure*}[t!]
  \centering
    \psfig{file=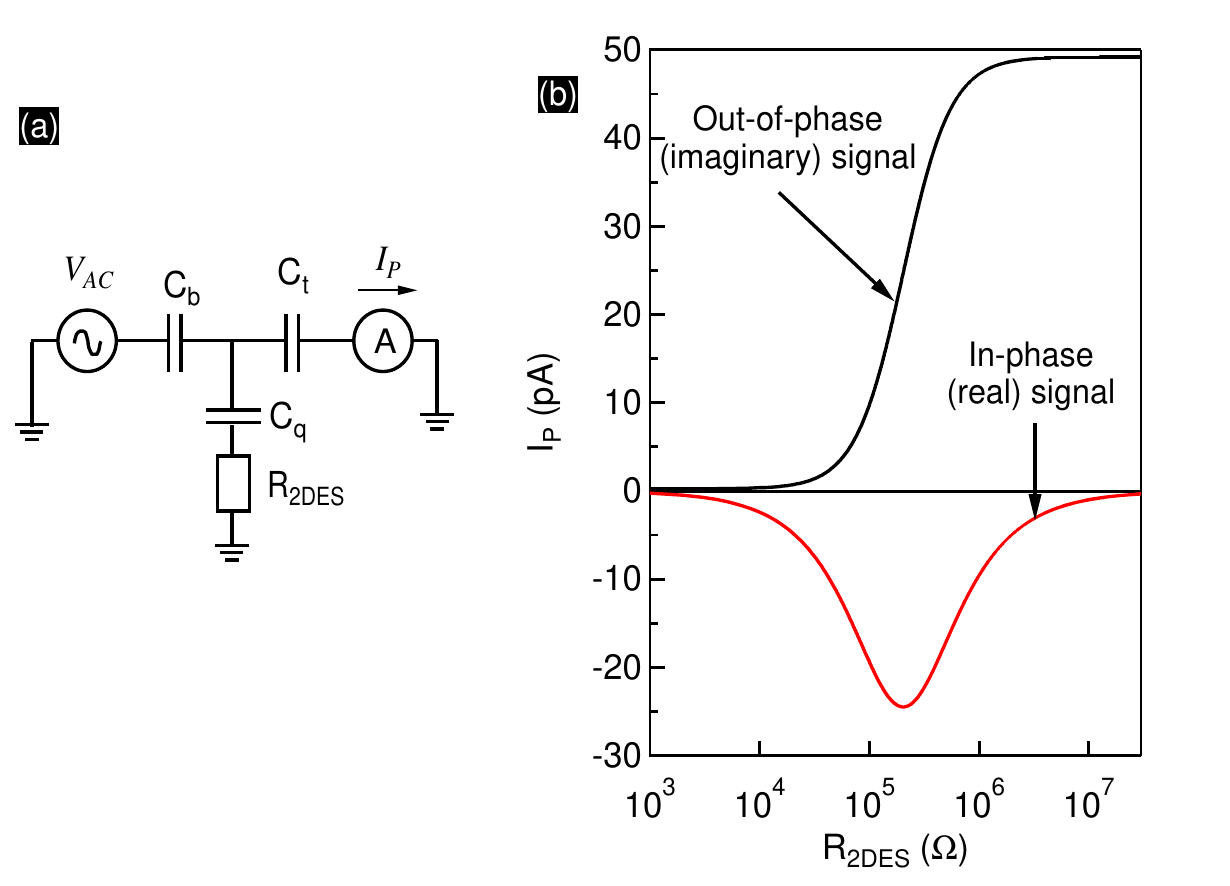, width=0.9\textwidth}
  \centering
  \caption{\label{transport} 
   (a) Lumped-element circuit schematic for our capacitance measurement set-up. The quantum capacitance and resistance of the 2DES are denoted as $C_{q}$ and $R_{2DES}$, respectively. We did not included parasitic capacitances in the model. (b) Penetrating current $I_{P}$ vs. $R_{2DES}$, calculated for $f=213$ Hz. The signal is complex with the Imaginary component carrying the bulk of the information regarding capacitances.   
  }
  \label{fig:transport}
\end{figure*}

\begin{figure*}[t!]
  \centering
    \psfig{file=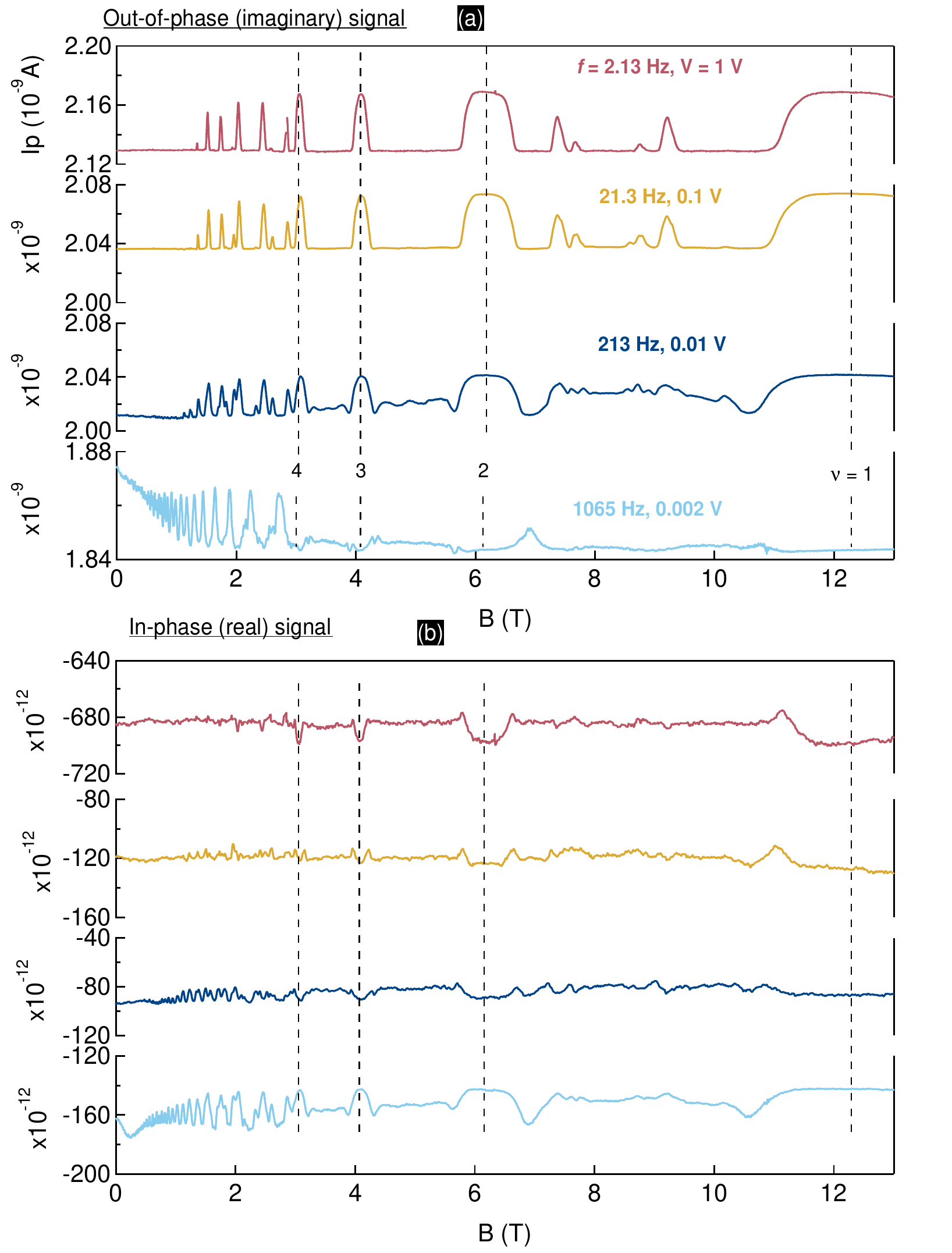, width=0.9\textwidth}
  \centering
  \caption{\label{transport} 
   (a) and (b) show the measured raw $I_{P}$ vs. $B$ (imaginary and real components) for four different frequencies: $2.13$, $21.3$, $213$, and $1065$ Hz; the corresponding values of applied $V_{AC}$ are also indicated in each trace. In the measurements we change the applied voltages in order to keep the product of the frequency and the voltage constant. Integer quantum Hall states are identified as marked by vertical dashed lines. 
  }
  \label{fig:transport}
\end{figure*}

\begin{figure*}[t!]
  \centering
    \psfig{file=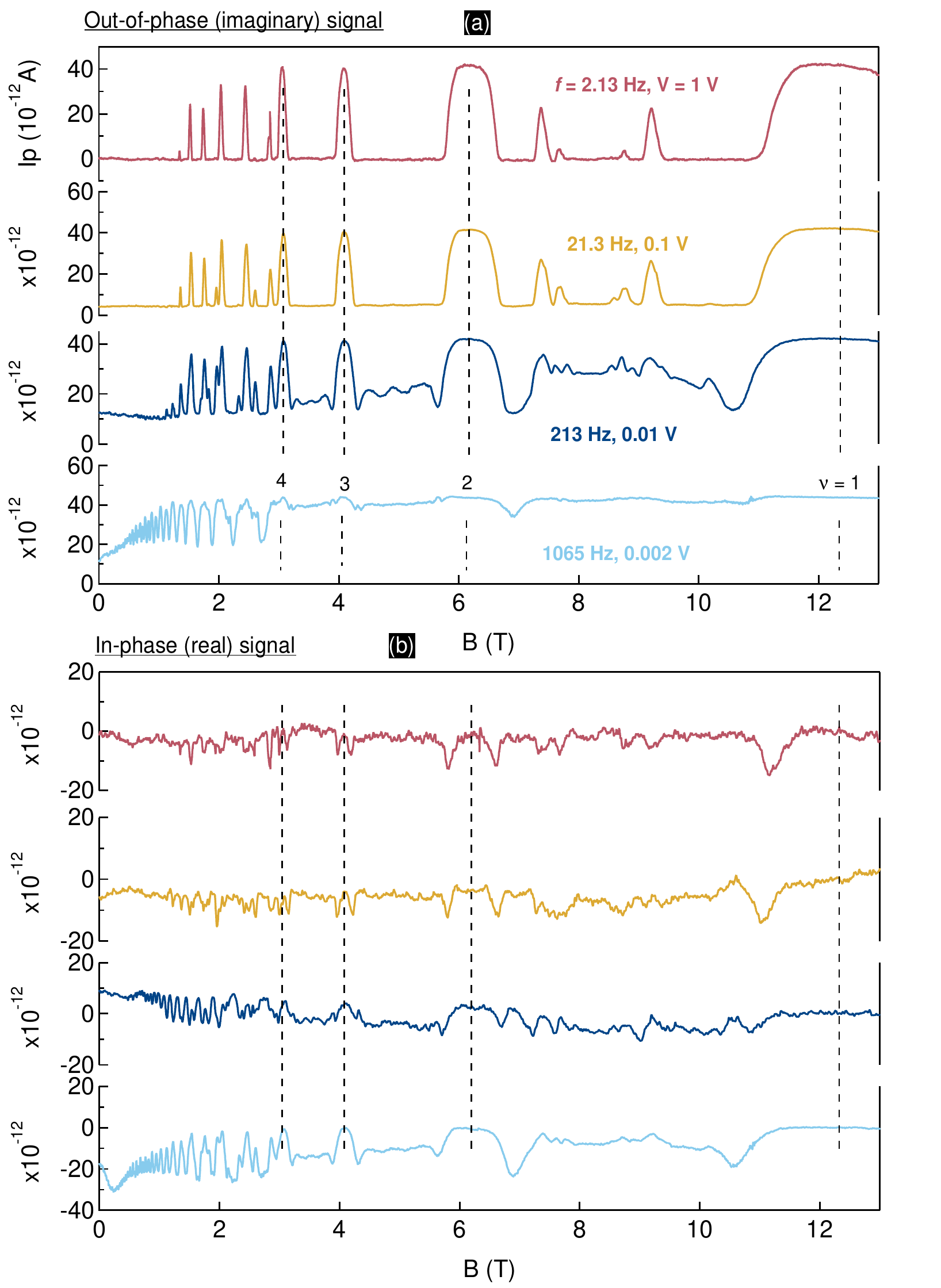, width=0.9\textwidth}
  \centering
  \caption{\label{transport} 
   (a) and (b) show the calibrated $I_{P}$ vs. $B$ (imaginary and real components) for four different frequencies: $2.13$, $21.3$, $213$, and $1065$ Hz; the corresponding values of $V_{AC}$ are also indicated in each trace.
  }
  \label{fig:transport}
\end{figure*}

\clearpage

\section{Additional $I_{P}$ data: other Landau level fillings and temperature dependence}

Figures S5(a-d) show $I_{P}$ for a wide range of filling factors $\nu=4-8$, at $T\simeq60$ mK. The bubble phases are denoted with vertical arrows pointing to the local maxima in $I_{P}$. We observe signatures of bubble phases in $I_{P}$ up to $\nu=7$.

The non-monotonic temperature dependence of $I_{P}$ shown in Figs. 3(b,c) in the main manuscript can also be observed when $T$ is fixed and magnetic field sweeps are taken. Figure S6(a) shows $I_{P}$ traces at different temperatures for $\nu$ ranging from $4$ to $5$. The bubble phases are marked by vertical dashed lines. Here we describe the temperature dependence near $\nu\simeq4.30$ and $4.73$. For $\nu\simeq4.30$, at the lowest temperatures, $I_{P}$ is high, and as the $T$ increases, $I_{P}$ reaches a minimum near $T\simeq134$ mK. At higher temperatures, $I_{P}$ saturates at a value smaller than the low temperature limit. A qualitatively similar behavior is observed for the bubble phase around $\nu\simeq4.73$. In Fig. S6(a) we can also observe the evolution of $I_{P}$ in the anisotropic phase region detected in magnetotransport, namely around $\nu\simeq4.50$. In the high-temperature limit, $I_{P}$ is close to $18$ pA for $\nu$ near to $4.50$. As $T$ is decreased, $I_{P}$ starts to decrease, initially at $\nu=4.50$, and then expanding to a wider range of $\nu$. In the low-temperature limit, $I_{P}$ is nearly constant at around $13$ pA for a range of $\nu$. Hence, the anisotropic states screen better than the high temperature liquid state.


\begin{figure*}[h!]
  \centering
    \psfig{file=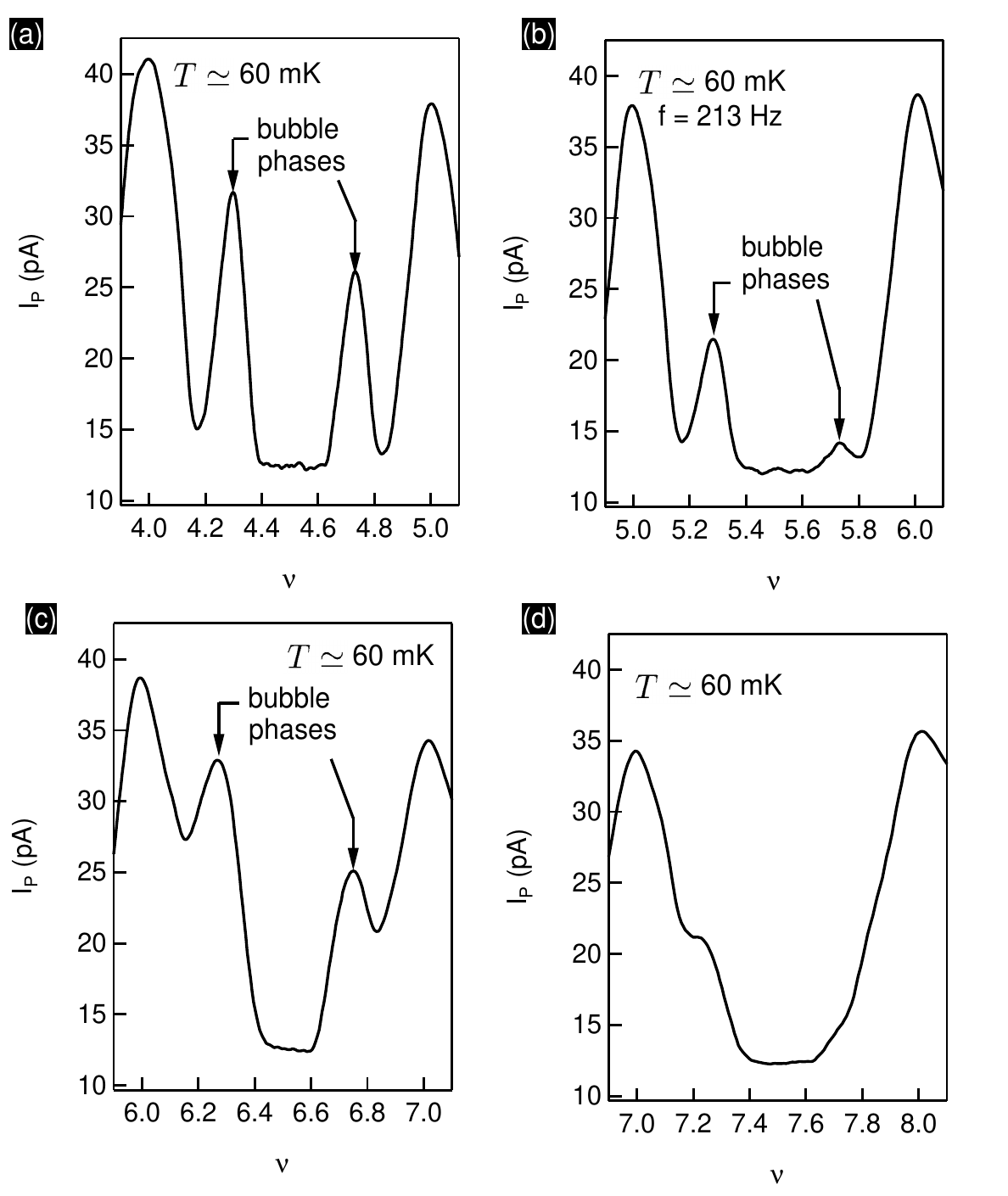, width=0.95\textwidth}
  \centering
  \caption{\label{transport} 
   $I_{P}$ vs. $\nu$ at $T\simeq60$ mK. The bubble phases are denoted with vertical arrows. 
  }
  \label{fig:transport}
\end{figure*}

\begin{figure*}[h!]
  \centering
    \psfig{file=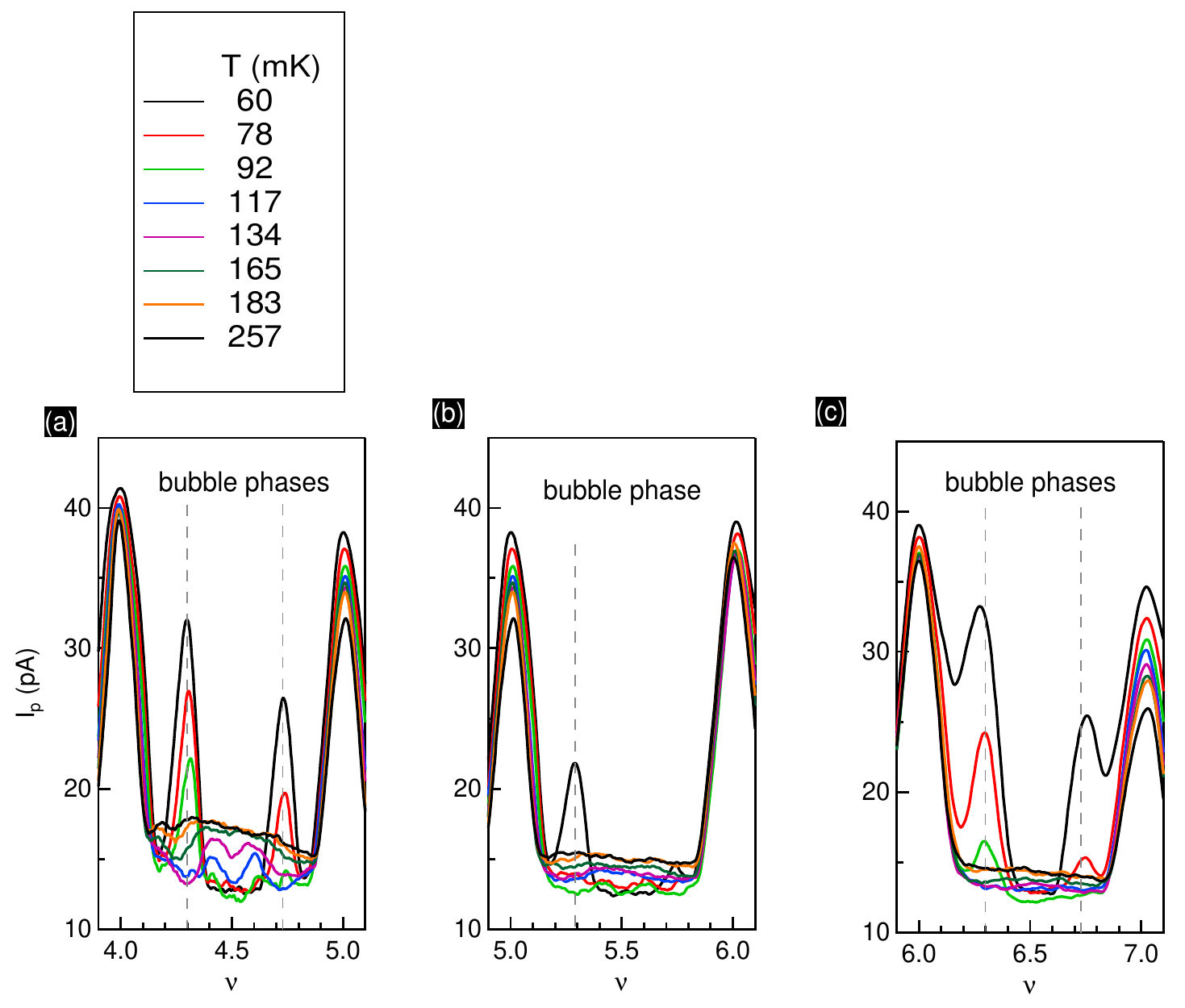, width=1\textwidth}
  \centering
  \caption{\label{transport} 
   $I_{P}$ vs. $\nu$ for different temperatures ($T$). The bubble phases are marked with dashed vertical lines.
  }
  \label{fig:transport}
\end{figure*}

\clearpage

\section{Frequency dependence of the $I_{P}$ data near the melting of bubble phases}

Figures S7(a-c) show $I_{P}$ vs. $T$ data at $\nu=4.30$ for three different frequencies (after performing the calibration outlined in the first section). The non-monotonic $I_{P}$ response and the value of the bubble phase melting temperature $T_{M}$ are frequency independent in the range from $21.3 - 1000$ Hz. We also wish to emphasize that the resistive (in-phase) part of the $I_{P}$ signal near $T_{M}$ for $f=21.3$ Hz is close to zero within the noise level of our measurements.

\clearpage

\begin{figure*}[t!]
  \centering
    \psfig{file=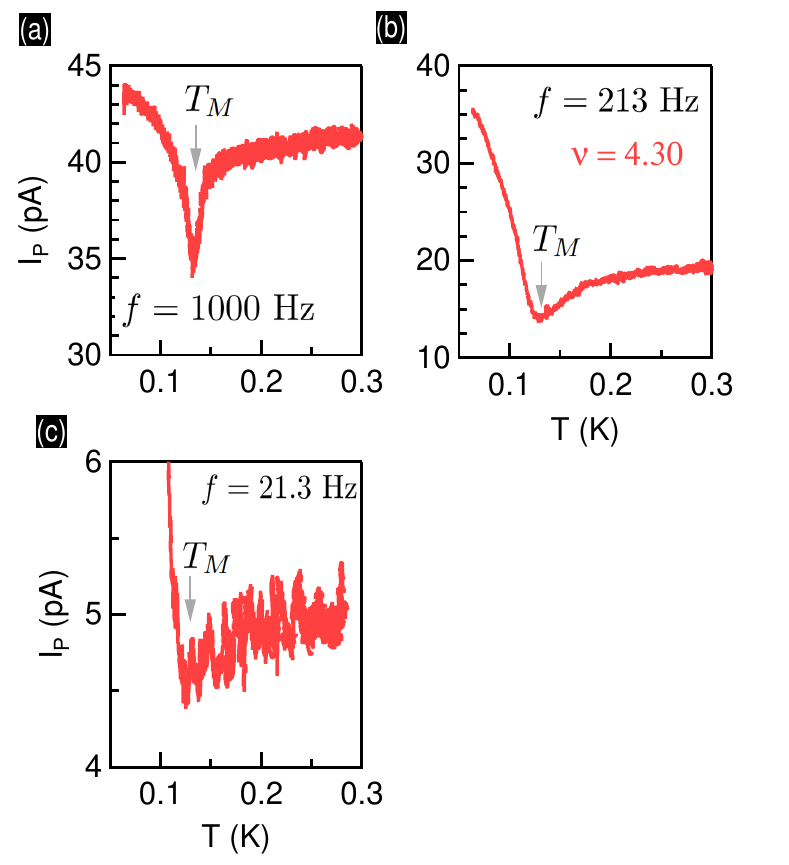, width=0.8\textwidth}
  \centering
  \caption{\label{transport} 
   $I_{P}$ vs. $T$ at $\nu=4.30$ for three different frequencies (a) $1000$, (b) $213$, and (c) $21.3$ Hz. Each trace is taken using a different value for $V_{AC}$: 0.002 V, (b) 0.01 V, and (c) 1 V.
  }
  \label{fig:transport}
\end{figure*}